# Verification of Short-Range Order and Its Impact on the Properties of the CrCoNi Medium Entropy Alloy

Ruopeng Zhang[1,2,†], Shiteng Zhao[1,2,†], Jun Ding[3], Yan Chong[1,2], Tao Jia[4], Colin Ophus[2], Mark Asta[1,3], Robert O. Ritchie[1,3] & Andrew M. Minor[1,2,*]

[1] *Department of Materials Science and Engineering, University of California, Berkeley, CA, USA.*

[2] *National Center for Electron Microscopy, Molecular Foundry, Lawrence Berkeley National Laboratory, Berkeley, CA, USA.*

[3] *Materials Sciences Division, Lawrence Berkeley National Laboratory, Berkeley, CA, USA*

[4] *Department of Physics, Stanford University, Stanford, California 94305, USA*

[†] *These authors contributed equally to this work.*

[*] *Correspondence to aminor@lbl.gov*

**Traditional metallic alloys are mixtures of elements where the atoms of minority species tend to distribute randomly if they are below their solubility limit, or lead to the formation of secondary phases if they are above it. Recently, the concept of medium/high entropy alloys (MEA/HEA) has expanded this view, as these materials are single-phase solid solutions of generally equiatomic mixtures of metallic elements that have been shown to display enhanced mechanical properties. However, the question has remained as to how random these solid solutions actually are, with the influence of chemical short-range order (SRO) suggested in computational simulations but not seen experimentally. Here we report the first direct observation of SRO in the CrCoNi MEA using high resolution and energy-filtered transmission electron microscopy. Increasing amounts of SRO give rise to both higher stacking fault energy and hardness. These discoveries suggest that the degree of chemical ordering at the nanometer scale can be tailored through thermomechanical processing, providing a new avenue for tuning the mechanical properties of MEA/HEAs.**

The attractive concept of using multiple principal elements in a single alloy system has received a tremendous amount of interest over the past decade[1–5]. This group of materials are usually termed as medium-entropy alloys (MEA) in ternary systems and high-entropy alloys (HEA) in quaternary or quinary systems, alluding to their high degree of configurational entropy. Among the increasing number of medium- to high- entropy alloy systems reported in the literature[6–10], the CrCoNi-based, face-centered-cubic (*fcc*) single-phase alloys exhibit an exceptional combination of mechanical properties including high strength, tensile ductility, fracture toughness and impact resistance[11]. Extensive studies have been documented on the deformation mechanisms in these alloys. Otto et al.[12] evaluated the temperature dependence of plasticity behavior in CrMnFeCoNi and observed a slip-twinning transition at reduced temperatures[12]. Gludovatz et al. later reported the outstanding fracture toughness of CrCoNiFeMn[4] and CrCoNi[13] at cryogenic temperatures and attributed this to a synergy of deformation mechanisms including a propensity for mechanical twinning[14]. Interestingly, computational work has suggested that the CrCoNi-based *fcc* single-phase alloys should have near-zero stacking fault energies (SFE, $\gamma_{SF}$)[15]; further calculations for random HEA and MEA solid solutions even suggest that these values could be negative at low temperatures[14,16–18]. These computational predictions do not agree quantitatively with measured values[19,20] ($\gamma_{SF\_CrCoNi}$ ~22 mJ/m$^2$ and $\gamma_{SF\_CrMnFeCoNi}$ ~30 mJ/m$^2$). Experimentally, the measured SFEs in MEAs and HEAs derived from separation of Shockley partial dislocations exhibit a wide distribution[21], indicating a strong dependence of $\gamma_{SF}$ on local atomic configuration. Ding *et al.*[22] showed that the SFE of CrCoNi MEA can be tailored over a wide range of values by tuning its local chemical order. The work highlights the potentially strong impact of chemical short-range order (SRO, defined as the deviation in the number of different chemical bond types from those corresponding to a random solid solution) on the mechanical properties of the MEA/HEAs. Later, Li *et al.*[23] demonstrated the ruggedness of the local energy landscape and how it raises activation barriers governing dislocation activities with molecular dynamics simulations. However, to date experimental evidence for the existence of such SRO has been limited to X-ray adsorption measurements[24] that are averaged over a relatively large volume of material. Indeed, further efforts are needed to characterize the degree and nature of ordering, its spatial extent, how both depend on thermal history in processing, and the associated effects on mechanical behavior. Here, we provide the first direct and quantitative visualization of the SRO structure, where we establish a direct effect of this SRO on the mechanical behavior of MEA/HEA materials. The results thus establish a novel design parameter to control the properties of MEA/HEAs that we expect could be applicable to the wide class of multi-principal-element alloys more generally.

To investigate the presence of chemical SRO, equiatomic CrCoNi alloy samples were arc-melted and homogenized at 1200 °C for 48 h. The alloy was then subjected to different thermal treatments: (1) water-quenched to room temperature (RT), or (2) aged at 1000 °C for 120 h followed by furnace cooling to RT. All the alloys were confirmed to be a single-phase *fcc* structure via X-ray diffraction and electron backscatter diffraction (EBSD) analysis. The microstructure and the degree of SRO were then characterized with a variety of Transmission Electron Microscope (TEM) imaging techniques. Diffraction contrast from SRO is inherently faint as compared to the long-range *fcc* lattice diffraction signal since it arises from the relatively minor differences in atomic scattering factors between the constituent elements. As a result, measurement of the relatively faint SRO diffraction signal has proven to be challenging. In order to enhance the signal-to-noise ratio of the diffraction contrast from SRO, we minimized the background noise from inelastic scattering by using a Zeiss LIBRA 200MC microscope, equipped with an in-column Ω energy filter and a 16-bit dynamic range camera. Energy-filtered diffraction patterns and dark-field (DF) images for the two heat treatment conditions are shown in Fig. 1. In the diffraction patterns (Figs. 1a-b), streaks along {111} directions between *fcc* Bragg spots are clearly observed in the aged sample. Dark field imaging taken with the objective aperture positioned in the center of the streaked region shown in Fig. 1b were used to directly image the SRO domains. While no DF contrast can be seen from the water-quenched samples (Fig. 1c), the aged sample (Fig. 1d) clearly reveals the nanoscale domains. Results from an intermediate heat treatment are shown in Supplementary Fig. 1 for comparison.

The diffuse scattering in the diffraction patterns and associated contrast in the dark-field images could arise from a combination of effects, including static and thermal displacement scattering and chemical SRO[25]. However, the fact that the water-quenched samples (Fig. 1c) show negligible contrast using the same imaging conditions, and the fact that the aged samples show enhanced streaking and DF contrast strongly supports the interpretation that these features are associated with chemical SRO. This suggests that the streaking contrast in the diffraction patterns is caused by the SRO domains that form a diffuse superlattice with a distortion of the local lattice. Specifically, the enhanced contrast in samples aged at higher temperature can be interpreted to be associated with the higher mobility of the atoms at these temperatures, which is able to evolve towards a lower free energy state with higher chemical SRO. Further evidence in support of this interpretation follows.

High-resolution TEM imaging (HRTEM) has been used to distinguish the difference between thermal- and static-displacement induced diffuse scattering in previous studies[25]. Figs. 1e-f shows a comparison of water-quenched and aged HRTEM images, where two regions in the aged sample show diffuse superlattice features along {111} planes as marked in Fig. 1f. In addition, the 2-D fast fourier transforms (FFT) of the HRTEM images (Figs. 1e, f insets), show a similar streaking intensity along the {111} g vectors. These observations provide clear evidence that the contrast in the real-space HRTEM images is associated directly with the diffuse intensity observed in the diffraction patterns. The features observed in the HRTEM images are qualitatively consistent with the type of order suggested in both the EXAFS[24] and previous Monte-Carlo simulations[22,26], both of which indicate that Cr-Cr pairs are strongly disfavored at nearest-neighbor distances such that the concentrations of Cr atoms in {111} planes could be expected to alternate from enhanced to depleted values along the various <111> directions. Such compositional variations are consistent with the alternating contrast along the <111> directions observed by HRTEM.

The combined evidence from diffraction contrast and HRTEM imaging is that the high-temperature aging leads to the formation of appreciable SRO in *fcc* CrCoNi MEAs. The size and shape of the SRO-enhanced domains can thus be evaluated through energy-filtered DF imaging. For example, Figures 2a-b present two DF images formed by using two different objective aperture positions as marked in Fig. 2c. While each DF image (Figs. 2a, b) shows mostly different sets of SRO enhanced domains that are preferentially scattering to different parts of reciprocal space, there are a number of domains that could be identified in both images (examples are marked by the arrows). The existence of the same domains in images formed by separate and non-parallel directions of SRO-generated streaking is evidence for a non-planar shape of the SRO-domains.

It is also possible to characterize the size distribution of the domains by assuming a shape (in this case we assume a spherical shape for simplicity) and applying a Gaussian template fitting algorithm[27] as demonstrated in the Methods section. This analysis generates an average diameter of the measured domains of 1.13 ± 0.43 nm, which would correspond to the 3$^{rd}$ to 4$^{th}$ atomic shells on the *fcc* lattice of CrCoNi MEA[16,19,28]. However, as the DF images in Figs. 1, 2 suggest, the domain boundaries are relatively diffuse, and there is no evidence of any specific shape that characterizes the SRO domains. Further evidence for the diffuse nature of the SRO domains could be obtained by conducting geometrical phase analysis (GPA) on drift-corrected high-resolution scanning transmission electron microscope (STEM) images[29]. The resulting strain maps are summarized in Supplementary Fig. 2. In the water-quenched sample, the fluctuation of local strain is minimal. However, in the 1000 °C aged sample, domain contrast similar in size to that found in the DF images could be identified, indicating a small yet locally ordered fluctuations in lattice distortions. The results are suggestive that the SRO may be associated with the changes in the static atomic displacements, which is of interest since lattice distortions are widely proposed to partially explain the mechanical properties of the CrCoNi MEA[11]. This result thus warrants further investigation. We note, however, that standard X-ray diffraction (XRD) experiments conducted on both water-quenched and 1000 °C aged samples show immeasurable changes in peak broadening for the two different thermal treatments (Supplementary Fig. 3), such that further investigations of the lattice distortions would likely require synchrotron measurements and lie beyond the scope of the present study.

It is known that the formation of SRO has a significant impact on dislocation plasticity, where an increasing degree of SRO tends to increase the planarity of dislocation slip[30–32]. To assess the effect in the CrCoNi alloy, dislocation analysis was conducted on bulk compressed samples and the results are summarized in Fig. 3. Specifically, a random distribution of dislocations was observed in the water-quenched sample, whereas a marked trend of localized planar configuration of dislocations was present in the 1000 °C aged sample with SRO (Figs. 3a, b). In the latter case, the leading dislocations also tend to form dislocation pairs, where the separation distance of two adjacent dislocations were significantly reduced (two examples were marked by the white arrows in Fig. 3b). One possible origin of planar slip in *fcc* materials is the Shockley partial dissociation of perfect dislocation cores, limiting the ability of cross slip. In the current study, however, the aged alloy possesses more compact dislocation cores than the quenched alloy while presenting planar slip. On the other hand, localized planar slip and leading dislocation pairs are usually correlated to the glide plane softening effect due to the local destruction of the SRO structure[30,32,33], where the initial dislocation motion interrupts the SRO atomic configuration to overcome the energy barrier associated with the creation of a diffuse-anti-phase boundary (DAPB). Subsequently, dislocations following the initial dislocation would experience a lower energy barrier by gliding on the same path and avoiding the DAPB energy barrier. The DAPB energy as a function of dislocation slip events has been assessed by Density Functional Theory (DFT) calculations based on the calculated SRO atomic configuration[22], supporting this theory on the origin of the planar dislocation slip (Supplementary Fig. 4).

The exceptional strength, ductility and toughness of CrCoNi MEA can be directly correlated with the SFE of the material[11]. Previous DFT-assisted Monte Carlo simulations predicted that the SFE of CrCoNi MEA could be highly tunable by varying the SRO.[22] While the SFE of MEA/HEAs has been experimentally probed previously via both weak-beam DF imaging[19] and diffraction contrast STEM (DC-STEM) analysis[21], the SFE has never been directly correlated to the degree of SRO. In the current study, the SFE was measured by DC-STEM analysis as the technique allows for imaging through thicker samples to minimize the sample surface effect. Figures 3c, d show examples of images where partial dislocations could be identified through "$g \cdot b$" analysis and their disassociation measured directly (detailed in Supplementary Fig. 5). The separation distance and the statistical results are summarized in Fig. 3e and Table 1. The detailed calculation of the stacking fault energy is elaborated in the Methods section, which results in the 1000 °C aged samples having an SFE of 23.33 ± 4.31 mJ/m$^2$, double the value of its water-quenched counterpart (8.18 ± 1.43 mJ/m$^2$). This measurement confirms that the SRO directly impacts the SFE and indicates that the SFE could by fine-tuned by controlling the ordering, exactly consistent with the DFT-assisted Monte Carlo simulations.[22] It is noteworthy that the SFEs of both water-quenched and 1000 °C aged samples are distributed within a range, but demonstrate a clear trend.

In order to quantify the impact of SRO on the MEA's mechanical properties, both nanoindentation tests and bulk tensile tests of the same alloys were performed. The measured nanoindentation hardness is 4.07 ± 0.23 GPa for the water-quenched sample and 4.37 ± 0.58 GPa for the 1000 °C aged sample. The results show that the MEA is heat treatable, similar to the bake hardened steels or some aluminum alloys. SRO also significantly affects the onset of plasticity which is manifested by the "pop-in" event[34] in the load *vs*. displacement curves in Figs. 4c,d. The first pop-in events of the SRO-aged sample are distributed more discretely and usually occur later (at a higher load) than the quenched sample. In addition, the displacement plateau that corresponds to the strain burst of a pop-in event is larger in the aged material, as detailed in Supplementary Fig. 6. The higher pop-in load and larger displacement plateau in the SRO-aged specimen indicates the presence of dislocation avalanches (sudden bursts of dislocation nucleation and propagation), providing another evidence of the SRO hardening and the subsequent glide plane softening caused by passage of the first few dislocations in the slip band. Bulk tensile tests confirmed the strengthening effect of SRO by showing an ~ 25 % increase of the yield strength (Table 1) as well as a dramatic change of the work hardening behavior. As demonstrated in Figs. 4g, h, the initial work hardening rate of the aged sample is as twice as much of its water-quenched counterpart, reinforcing that the hardening is caused by the SRO domains. Traditionally, the formation of SRO in alloys causes planar dislocation slip and deformation localization[30,35–37]. In some cases, the deformation localization affects the alloys' ductility and toughness, whereas in the current study, the formation of SRO has little effect on the overall ductility of the MEA alloy. Deformation twinning is reported to explain the exceptional ductility of the CrCoNi alloy[11,14], in which nano-twinning delays deformation localization. Though lacking direct evidence, considering the similar work hardening behavior at the later stage of the deformation of both the SRO-aged and the water-quenched samples, we speculate that the exceptional strength and toughness of CrCoNi MEA arises in part from this unique combination of SRO hardening and twin-induced deformation at later stages. However, further systematic analysis is required to fully understand any potential effect of SRO on the deformation twinning.

In addition to the effect of SRO on plastic behavior, it also, in theory, should affect elastic properties as the local bonding environments are significantly altered from the perfect random solid solution. A simple rule-of-mixtures would predict a Young's modulus of ~229 GPa for equiatomic CrCoNi[38]. However, the nanoindentation modulus (reduced modulus) of the water-quenched sample is measured to be 181.76 ± 13.37 GPa, 18.1% smaller than that of the 1000 °C aged sample (214.79 ± 18.49 GPa). In contrast, the global Young's moduli of the bulk materials were determined by ultrasonic pulse-echo technique where the longitudinal and transverse sound speeds are measured to calculate elastic modulus. The measured global Young's modulus of the water-quenched and the aged samples are 229.93 GPa and 230.99 GPa, respectively (other measured elastic properties are listed in Table 1). The discrepancy between the locally-measured modulus by nanoindentation and the bulk-scale modulus measured acoustically may result from the limited size (~1 nm) of SRO clusters. The local measurement of modulus by nanoindentation is sensitive to the homogeneity of the distribution of the SRO clusters. However, the wavelength of the ultrasonic acoustic waves used to measure global modulus is orders of magnitude longer than the size of the SRO. Therefore, the measurement is averaged over a much larger volume and is insensitive to the degree of SRO.

As an emerging class of structural materials, MEA/HEAs possess a desirable combination of mechanical properties for structural applications[11,39,40]. While the concept of MEA/HEAs is based on production of a single-phase solid solution, there has long been a question about how well-mixed the solid solutions are [4,6,11,24,41–45]. Here, we directly imaged the local ordering and showed how the deformation behavior of MEAs are directly correlated with the degree of SRO. Annealing the MEA to promote SRO led to an increase in hardness, a doubling of the SFE and a subsequent increase in planar slip. The phenomenon of planar slip has been widely observed in many systems[14,19,20,46,47], which can presumably also be linked to the degree of SRO in these systems. Due to its impact on the mechanical properties, the degree of SRO is a critical feature that should be considered in the materials design phase. Directly tailoring the SRO microstructure on an atomic level therefore provides another dimension for controlling the structure-property relationship of advanced materials.

## Methods

**Materials and sample preparation.** The raw ingot of the equiatomic CrCoNi MEA was argon arc double melted and then cut into smaller samples. The samples were then divided into two groups and underwent different thermal treatments, (1) homogenized at 1200 °C for 48 h then water quenched to RT (uniform texture, grain size ~800 µm as determined by EBSD), or (2) homogenized at 1200 °C for 48 h then aged at 1000 °C for 120 h followed by furnace cooling (uniform texture, ~1000 µm as determined by EBSD). Samples for dislocation analysis were further deformed by conducting bulk compression tests on an MTS Criterion (Model 43) system to introduce dislocation plasticity. The final strain was 6% with a strain rate of $1 \times 10^{-3}$. The samples were then sliced and thinned by mechanical polishing. Electron-transparent samples for TEM observation were prepared with a Fischione twin-jet Electropolisher using a solution of 70% methanol, 20% glycerol and 10% perchloric acid at -20 °C. The samples for nanoindentation tests were prepared by mono-side electrochemical polishing with aforementioned solution and parameters.

**Energy-filtered dark-field TEM imaging and SRO domain recognition.** TEM samples of different heat treatments were used for observation. A Zeiss LIBRA 200MC microscope, equipped with an in-column Ω energy filter, was used to take both diffraction patterns and dark-field images. It is necessary to consider the impact on the resolution from the objective aperture, which could be estimated by the Airy disk radius using the following equation[48],

$$r_{Airy} \approx \frac{1.2\,\lambda f}{D},$$

where $\lambda$ is the electron wavelength (0.02507 Å for 200 kV TEM), $f$ is the focal length of the objective lens (~ 3 mm for the Zeiss Libra) and $D$ is the diameter of the objective aperture (25 μm aperture used in the current study). For the experimental setup used in the current study, the size of the aperture Airy disk is 3.61 Å, which is below the size of the observed SRO domains. A 5-eV energy slit was deployed to select the zero-loss peak and eliminate the contrast from inelastic scattering. A Gatan US1000 CCD camera was used to acquire the diffraction patterns and DF images. Prior to the data analysis, the energy-filtered DF images were filtered by a dark reference subtraction. According to the energy-filtered DF image shown in Figs. 1, 2, there is no observable directional tendency of the domains. Therefore, we assumed a circular kernel signal from the domains for our analysis. SRO-enhanced domains were identified and measured through Gaussian template fitting, where 2-D convolutions with the DF image were conducted using a list of differently sized 2-D Gaussian templates (with different values of standard deviation[27]). The stack of result images was further analyzed through a circular Hough transform to identify all signal peaks. The intensity cutoff was set according to the best fit result. Overlapping entities were deleted to ensure an accurate size measurement. Details of the algorithm are described below.

1. The standard deviation range of the Gaussian templates was set to 3 to 40 pixels (with a 0.1 interval) based on the pixel size of the DF image (0.056 nm/pixel).

2. 2-D Gaussian kernels with the same resolution of the DF images were constructed. The radius assumed for the SRO-enhanced domains was set to 1.3*sigma to best match the contrast observed in the DF image.

3. To suppress the background noise during convolution, each Gaussian kernel was normalized by a larger Gaussian function to give a zero summation, using the expression:

$$G_{2D}(x,y,\sigma) = \frac{1}{2\pi\sigma^2} e^{-\frac{x^2+y^2}{2\sigma^2}} - \frac{1}{2\pi(1.5\sigma)^2} e^{-\frac{x^2+y^2}{2(1.5\sigma)^2}}, \quad (1)$$

where $G_{2D}$ is the normalized 2-D Gaussian template, σ is the varying standard deviation of the differently sized kernels, and *x* and *y* are the 2D coordinates from the kernel origin.

4. The domain signals in the DF image were identified by a 2-D Gaussian Hough transform. For each kernel in the list, a 2-D convolution between the DF image and the kernel would be performed using:

$$y[m,n] = \sum_j \sum_i x[i,j] \cdot h[m-i, n-j], \qquad (2)$$

where $x$ is the DF image, $h$ is the kernel and $y$ is the result of the convolution.

5. After each convolution, pixels of the convolution result are compared to a data storing array; if the current pixel has a higher signal, the corresponding value would be updated in the data storing array. Another similarly sized array was used to store the associated kernel size of the highest signal.

6. After the iterations, the pixel values were first filtered by the domain diameter range and the peak signal cutoff. Then the local peaks in the result array were identified if a pixel has higher value than all of its eight neighbor pixels.

7. The identified peaks were ordered and checked in a "brightest to dimmest" manner according to their pixel value. If dimmer peaks appear in the radius of a brighter peak, they would be deleted. This process is to eliminate overlapping entities.

8. The remaining peaks were treated as identified domains.

9. The diameter cutoff below 0.7 nm is set manually as this is already the size of the 1st nearest-neighbor shell of atoms in the MEA lattice, we do not regard anything below this value as SRO domain.

A manual sampling was carried out to estimate the domain sizes and gain a reference for the optimization of parameters. Two critical parameters that would impact the identification are the minimal signal cutoff and the domain diameter range. The optimization process was conducted according to the best fitting results. In the case of a high signal cutoff or a narrow diameter range, the algorithm will miss some of the major contrast, whereas, in the case of a low signal cutoff or a wide diameter range, the algorithm will pick up lots of small intensity fluctuations that are from camera noise. It is worth mentioning the limitations to the domain recognition algorithm. Specifically, the assumption that the domains are spherical is for simplification, but the shapes of the domains vary. Parallel attempts of using a threshold segmentation algorithm involved much more subjectivity and yielded unreasonable results. However, the purpose of the analysis is to provide an estimated size distribution of the SRO domains, for which the current analysis is sufficient until large scale atomic imaging studies could provide similar statistics.

**X-ray diffraction (XRD) experiments.** The XRD experiments was performed *ex situ* with a PANalytical XPert diffractometer on water-quenched and 1000 °C aged samples, respectively. The scan range (2θ) was set to 42 – 54° to include the (111) and the (200) peaks. The angle resolution was set to 0.005° with a 0.8s integration time to ensure an accurate measurement of the lattice constants.

**High-resolution STEM (HRSTEM) imaging and geometrical phase analysis (GPA).** HRSTEM imaging of water-quenched and 1000 °C aged samples were conducted on the double-corrected TEAM I microscope (operated at 300 kV) at the National Center for Electron Microscopy (NCEM), Lawrence Berkeley National Laboratory. Drift correction was conducted with the methods developed by Ophus et al.[29] to eliminate the artifacts from beam scan jittering. FRWRtools plugin for Gatan Digital Micrograph software were used for the following GPA analysis. Averaged fast-fourier transforms were used as strain templates. The real-space resolution was set to 1.5 nm to achieve a relatively accurate measurement in reciprocal space.

**STEM EDS measurements.** Quantitative energy dispersive X-ray mapping (EDS) was conducted on both the water-quenched samples and aged samples using the TitanX microscope with a quad EDS detector. No chemical segregation was observed; results are summarized in Supplementary Fig. 7. The lack of any visible chemical segregation via EDS analysis in the aged samples is consistent with the high-resolution STEM observation presented in Supplementary Fig. 2, where there is no obvious Z-contrast difference despite different degrees of local lattice distortion. Previous theoretical studies[22,23] revealed that the SRO in the CrCoNi MEA is in the range of several nearest neighbor distances and that the driving force for the formation of SRO is to avoid certain types of bonding. Combined with the observation presented in the current study, we can conclude that it is not necessary for the SRO structure to possess a strong chemical segregation. Further verification using atomic-resolution EDS or electron energy loss spectroscopy (EELS) could provide valuable insights revealing the atomic structure of SRO clusters.

**Dislocation analysis.** TEM dislocation analysis was conducted on both the aged and water-quenched samples after 6% compressive deformation. TEM observations were conducted on the Zeiss LIBRA 200MC (operating at 200 kV) at NCEM. Low angle annular dark-field DC-STEM images[49,50] for partial dislocation "$g \cdot b$" and SFE measurements were acquired on the TEAM I microscope. The measured partial dislocation separation was further calibrated by conducting a g(3g) weak-beam dark-field imaging and calculating the actual partial separation from the observed values[19,51,52] The SFEs were calculated according to the following equation[19,53,54],

$$SFE = \frac{G b_p^2}{8\pi \cdot d}\left(\frac{2-\nu}{1-\nu}\right)\left(1 - \frac{2\cdot\nu\cdot\cos(2\beta)}{2-\nu}\right) , \qquad (3)$$

where $G$ is the shear modulus of CrCoNi MEA (determined by the ultrasonic pulse-echo measurement), $b_p$ is the magnitude of the Burgers vector of partial dislocations (~0.146 nm), $d$ is the measured separation of partial dislocations, $\nu$ is the Poisson's ratio (determined by the ultrasonic pulse-echo measurement), and $\beta$ is the angle between the perfect dislocation Burgers vector and the dislocation line. For both 1000 °C aged samples and water-quenched samples, 50 individual measurements were conducted on more than 10 partial pairs from relatively thick regions to avoid any surface effects. Associated ± standard deviations were calculated to ensure accurate and representative results.

**Nanoindentation experiments.** Nanoindentation tests were conducted on a Bruker Ti 950 TriboIndenter instrument with a 1-µm Berkovich tip. The peak load was set to 1000 µN. The analysis was conducted with a calibrated area function of the tip. The water-quenched and 1000 °C aged samples were electrochemically polished on one side with a solution of 70% methanol, 20% glycerol and 10% perchloric acid at -20 °C. A 10 × 10 grid of indents covering an area of 1 mm × 1 mm was set to conduct the test for each sample. No strong texture was observed by post-mortem EBSD. All quantitative parameters were averaged among the 100 indents with associated ± standard deviations.

**Bulk mechanical tests.** Bulk tensile tests were carried out on an MTS Criterion (Model 43) system. A Sony A7R Mark II camera were used to record images for Digital Image Correlation (DIC). A copy of Vic-2D Image Correlation software was utilized to conduct the DIC analysis. Due to the limited amount of material, the dimension of the gauge section of both water-quenched and 1000 °C aged samples was set to 5.1 mm × 0.8 mm × 1.6 mm. Specially designed sample grippers were utilized to conduct the tensile test. Sample surfaces were mechanically polished and sparkle-sprayed prior to the tests. The strain was extracted from the DIC Von-Mises strain data using 'virtual extensometers' mode and averaged 3 virtual extensometers along the gage length.

**Ultrasonic pulse-echo measurement.** An Olympus 38DL Plus thickness gauge with a Model 5072PR pulser/receiver module was used to measure the speed of the shear velocity and the longitudinal velocity. The Poisson's ratio, Young's modulus and the shear modulus were calculated with the following equations,

$$\nu = \frac{1-2(V_T/V_L)^2}{2-2(V_T/V_L)^2}, \tag{4}$$

$$E = \frac{V_L^2 \rho (1+\nu)(1-2\nu)}{1-\nu}, \tag{5}$$

$$G = V_L^2 \rho, \tag{6}$$

where $\nu$ is the Poisson's ratio, $V_T$ is the shear velocity, $V_L$ is the longitudinal velocity, $E$ is the Young's modulus, $G$ is the shear modulus and $\rho$ is the density of the materials, which is estimated with the following equation,

$$\rho = \frac{4 m_{aAvg}}{V_{cell}}, \tag{7}$$

where $m_{aAvg}$ is the averaged atomic mass of Cr, Co and Ni, $V_{cell}$ is the volume of a *fcc* unit cell calculated with the lattice constants derived from the XRD results.

**Nonproportional number of local atomic pairs.** To quantify the atom distribution at the alternating sites along <110> directions, we defined the nonproportional number of local atomic pairs around an atomic species for the 4$^{th}$ nearest-neighbor shell as detailed in a previous study[22] (corresponding to the second atom sites along the surrounding <110> directions),

$$\Delta \delta_{ij}^4 = N_{0,ij}^4 - N_{ij}^4, \tag{8}$$

where $N_{ij}^4$ is the actual number of pairs between atoms of type j and i in the 4$^{th}$ shell and $N_{0,ij}^4$ is the number of pairs that are proportional to the corresponding concentrations. Assuming entirely random atom distribution, $\Delta \delta_{ij}^4 = 0$, a positive value means a tendency of avoiding such a pair and a negative value means a tendency of forming such a pair.

**Diffuse anti-phase boundary energy.** The diffuse anti-phase boundary energy as the function of dislocation slip events was calculated via density functional theory using an "aged" atomic model reported in a previous literature[22], which has a similar SFE as the 1000 °C aged samples. Excess energy was calculated after each successive slip was introduced into the system.

## Data Availability

The data that support the findings of this study are available from the corresponding author upon reasonable request.

## Acknowledgements

This work was primarily supported by the Director, Office of Science, Office of Basic Energy Sciences, Materials Sciences and Engineering Division, of the U.S. Department of Energy under Contract No. DE-AC02-05-CH11231 within the Mechanical Behavior of Materials (KC 13) program at the Lawrence Berkeley National Laboratory. R.Z., S.Z. and Y.C. acknowledge support from the U.S. Office of Naval Research. Work at the Molecular Foundry was supported by the Office of Science, Office of Basic Energy Sciences, of the U.S. Department of Energy under Contract No. DE-AC02-05CH11231. Work at the Stanford Nano Shared Facilities (SNSF) is supported by the National Science Foundation under award ECCS-1542152. We thank Prof. Evan Ma at Johns Hopkins University for providing a 600 °C aged alloy.

## Author contributions

R.Z., S.Z., M.A., R.O.R. and A.M.M. conceived of the project.; R.Z. and S.Z. conducted the energy-filtered TEM imaging and dislocation analysis; C.O. and R.Z. developed and optimized the domain recognition algorithm; S. Z. conducted the nanoindentation tests; R.Z, S.Z and C.Y. conducted the tensile tests. J.D. conducted the DFT simulations. T.J. Conducted the XRD experiments. R.Z., S.Z., R.O R., M.A. and A.M.M. prepared the manuscript, which was reviewed and edited by all authors. Project administration, supervision, and funding acquisition was performed by R.O.R., M.A. and A.M.M.

## Author statements

Reprints and permissions information are available at www.nature.com/reprints. The authors declare no competing interests, financial or otherwise. Readers are welcome to comment on the online version of the paper.

## Data availability

The data that support the findings of this study are available from the corresponding author upon reasonable request.

## Correspondence

Correspondence and requests for materials should be addressed to aminor@berkeley.edu.

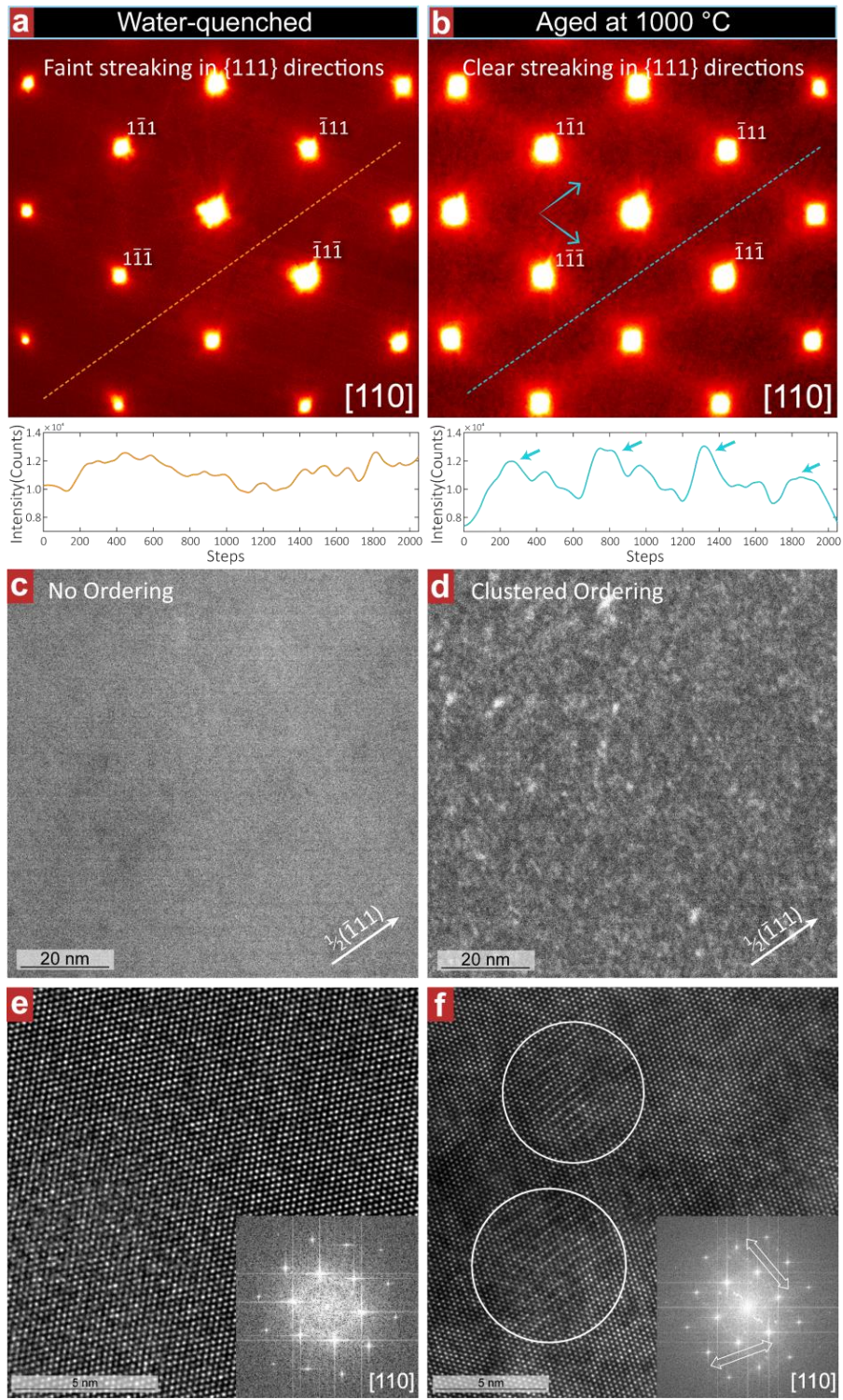

**Figure 1:** Energy-filtered TEM diffraction patterns, DF images formed with "diffuse superlattice" streaks and the associated high-resolution TEM images. (a-b), energy-filtered diffraction patterns taken from water-quenched and 1000 °C aged samples, respectively. The contrast is reversed and pseudo-colored for better visibility. The line plots of intensity show the periodic intensity of the "diffuse superlattice" streaks. (c-d), energy-filtered DF images taken from water-quenched and

1000 °C aged samples, respectively. The aperture positions are marked by the *g* vectors. (e), a typical high-resolution TEM image and the associated FFT image of a water-quenched sample. (f), a typical high-resolution TEM image and the associated FFT image of a 1000 °C aged sample. The features suggesting a superlattice are marked by the white circles and the associated streaking along the {111} directions is marked by the white arrows in the FFT image.

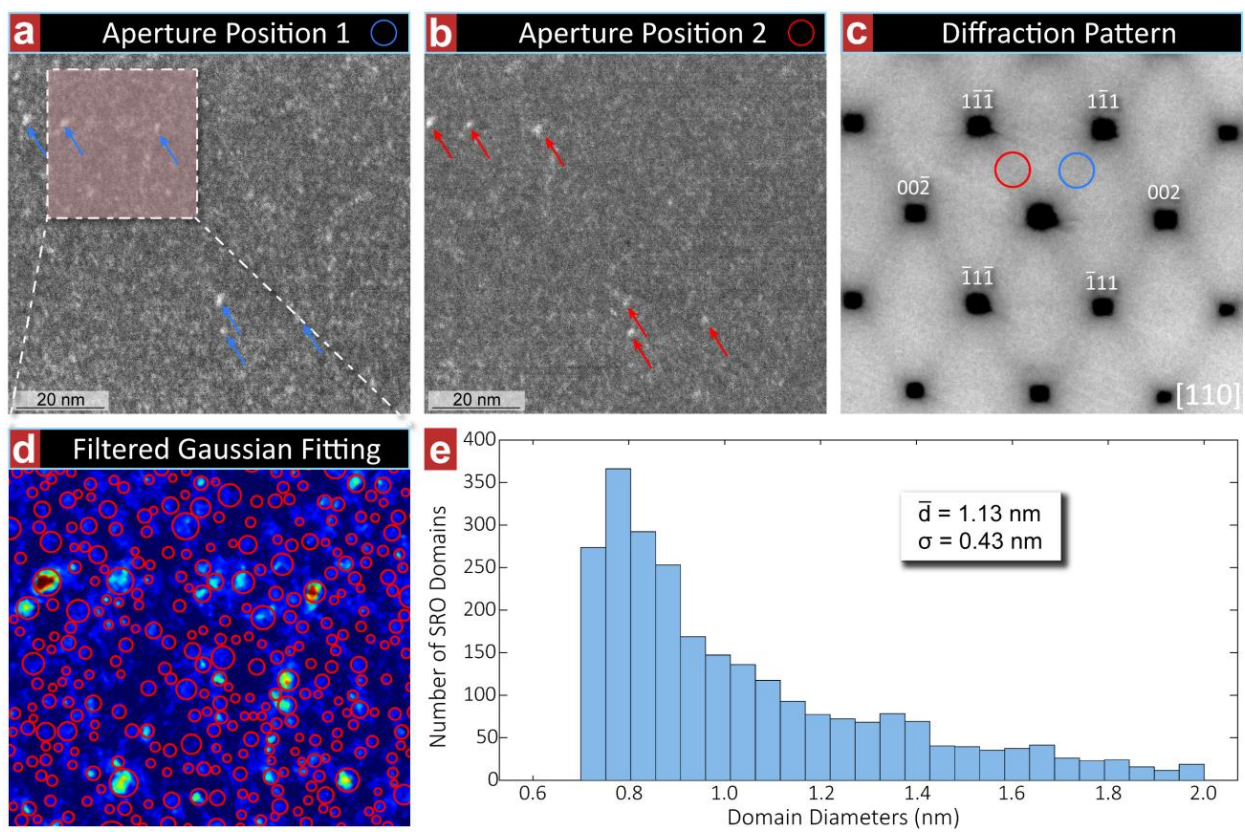

**Figure 2:** Evidence of the 3-dimensional structure of the domains and the size distribution of them. (a-b), energy-filtered DF images from different "diffuse" superlattice peaks; examples showing the same domain contrast are marked with the arrows. (c), energy-filtered diffraction patterns of the region of interest; the red and blue circles indicate the DF imaging conditions of a and b. The contrast is reversed for better visibility. (d), the enlarged DF image with identified SRO domains marked by the red circles. The DF image is pseudo-colored for better visibility. (e), the histogram of identified domain diameters. The average value and the standard deviation are listed in the box.

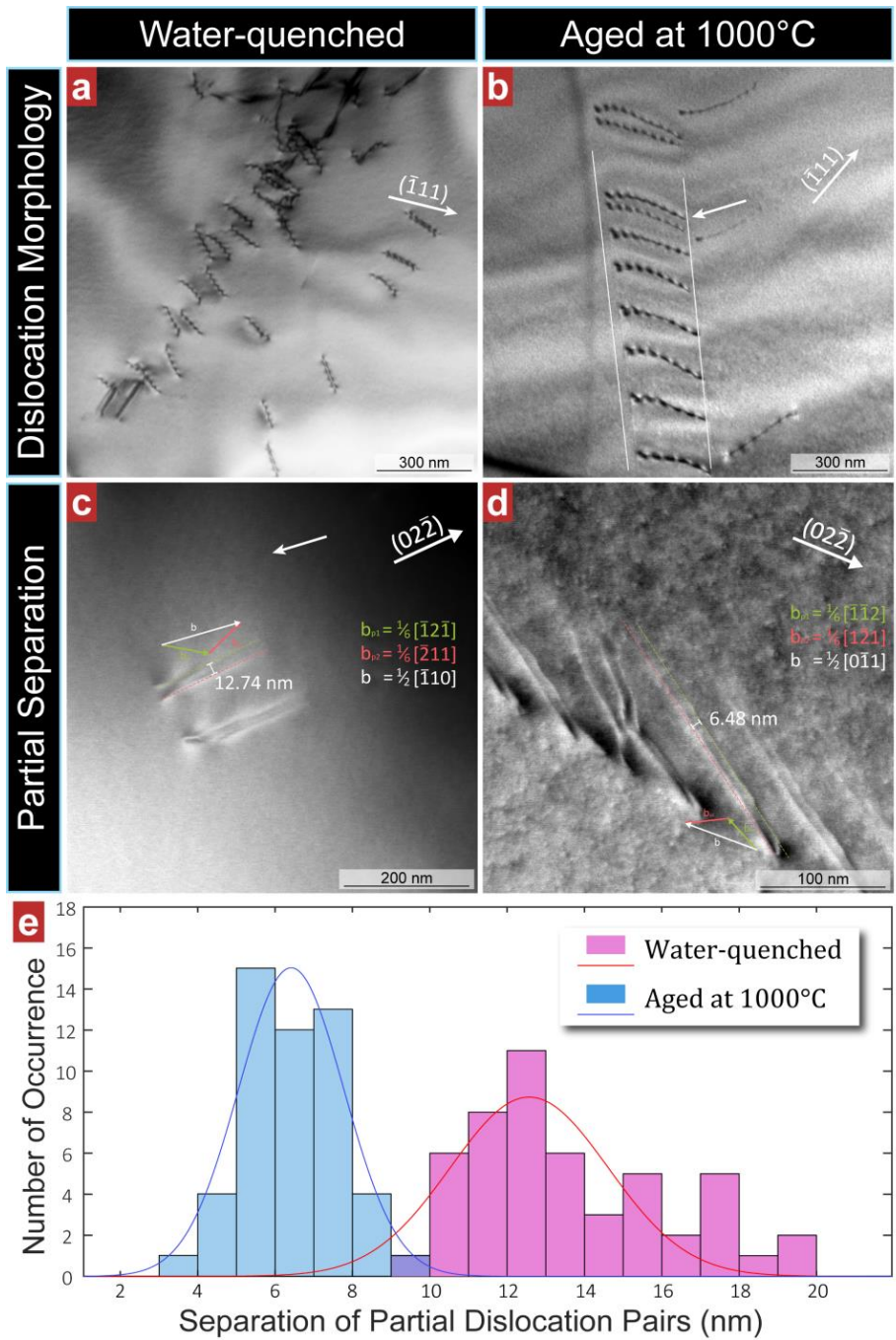

**Figure 3:** Dislocation analysis of both water-quenched and 1000 °C aged samples. (a), the two-beam bright-field (BF) image showing representative wavy configuration of dislocations in the water-quenched sample. (b), the two-beam BF image showing representative planar configuration of dislocations in the 1000 °C aged sample. The leading dislocation pairs are marked by the white arrows. (c), (d), LAADF images showing dislocation dissociations in water-quenched and 1000 °C aged samples, respectively; the Burgers vector relations are demonstrated in the figures; The detailed "$g \cdot b$" analysis is summarized in Supplementary Fig. 4. (e), distribution of the measured separation of partial dislocation pairs from both water-quenched and 1000 °C aged samples, respectively. The results of numerical analysis are summarized in Table 1.

| Water-quenched | Aged at 1000 °C |
|---|---|
| Load-Depth Curves of Nanoindentations 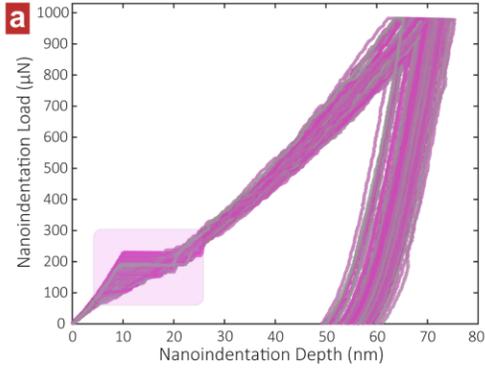 | Load-Depth Curves of Nanoindentations 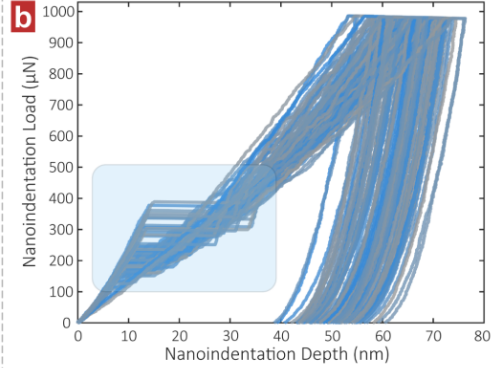 |
| Distribution of Pop-in Events 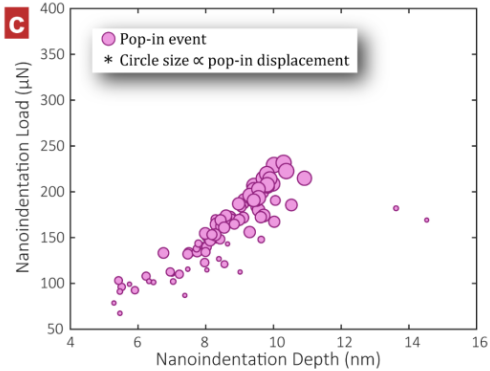 | Distribution of Pop-in Events 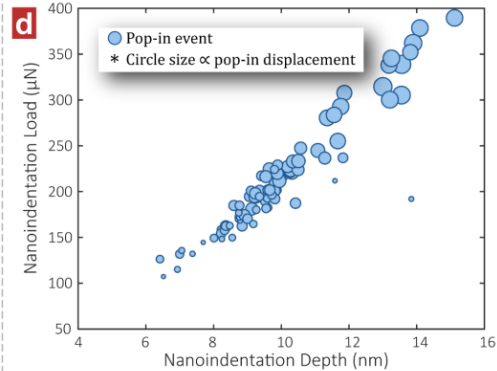 |
| True Stress Strain Curves 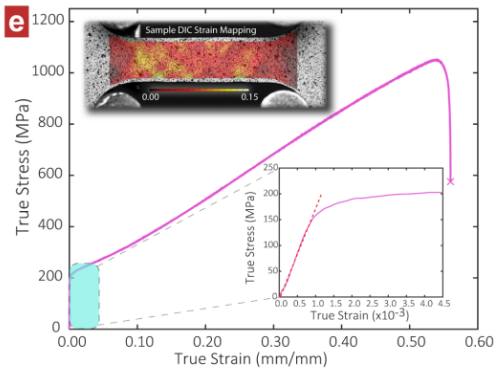 | True Stress Strain Curves 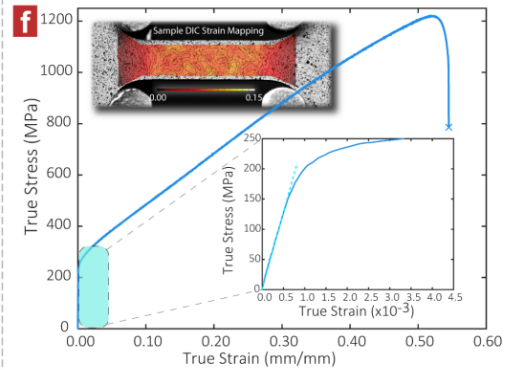 |
| Work Hardening Rate 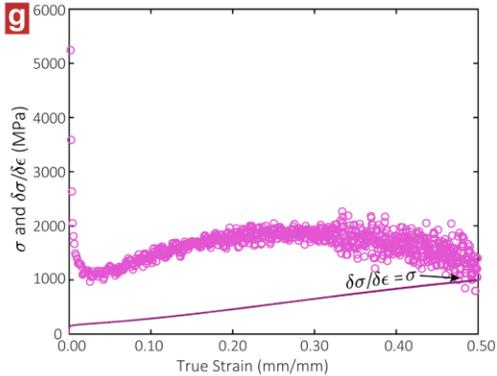 | Work Hardening Rate 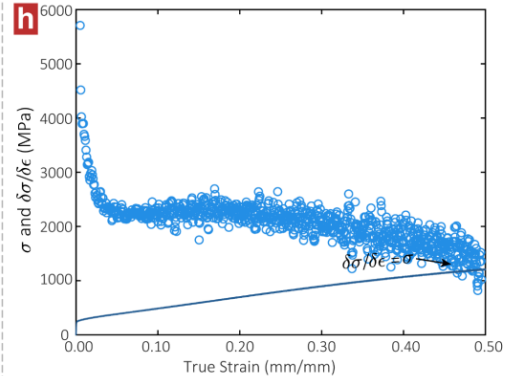 |

**Figure 4:** Comparison of mechanical properties from nanoindentation and bulk tensile tests. (a), (b), Load-depth curves from a 10 × 10 grid of nanoindentations separated by 10 μm from each other, from the water-quenched sample and 1000 °C aged samples, respectively. Pop-in analysis from these same tests are provided for the (c) water-quenched and (d) 1000 °C aged samples. The circles depict the depth and load of where the pop-in events occur. The sizes of the circles are proportional to the total pop-in displacement. (e), (f), results of tensile tests from the water-quenched sample and 1000 °C aged samples, respectively. Insets are the elastic portions of the curves and a sample image of the strain distribution during the elastic loading, as determined by DIC. (g), (h), work hardening rate derived from the true stress-strain curves of the water-quenched and the 1000 °C aged samples, respectively. True stress vs. true strain data from the same tests, respectively, are also displayed for comparison on (g), (h). The results of numerical analysis from these tests are summarized in Table 1.

**Table 1:** Statistical results of the SFE measurements and the nanoindentation tests.

|  |  | Water-Quenched | Aged at 1000˚C |
|---|---|---|---|
| **Elastic Properties** | **Poisson's Ratio** | 0.29 | 0.28 |
|  | **Young's Modulus (GPa)** | 229.9 | 231.0 |
|  | **Shear Modulus (GPa)** | 89.1 | 90.2 |
| **Yield Strength** | **0.2% Offset Yield Strength (MPa)** | 205 | 255 |
| **Dislocation Dissociation** | **Partial Separation, (nm)** | 13.59 ± 2.64 | 6.44 ± 1.19 |
|  | **SFE (mJ/m2)** | 8.18 ± 1.43 | 23.33 ± 4.31 |
| **Nanoindentation** | **Reduced Modulus (GPa)** | 181.76 ± 13.37 | 214.79 ± 18.49 |
|  | **Indentation Hardness (GPa)** | 4.07 ± 0.23 | 4.37 ± 0.58 |
|  | **Pop-in Load (µN)** | 164.52 ± 42.06 | 194.37 ± 36.06 |
|  | **Pop-in Starting Displacement (nm)** | 8.81 ± 1.04 | 9.40 ± 1.13 |

# Supplementary Information

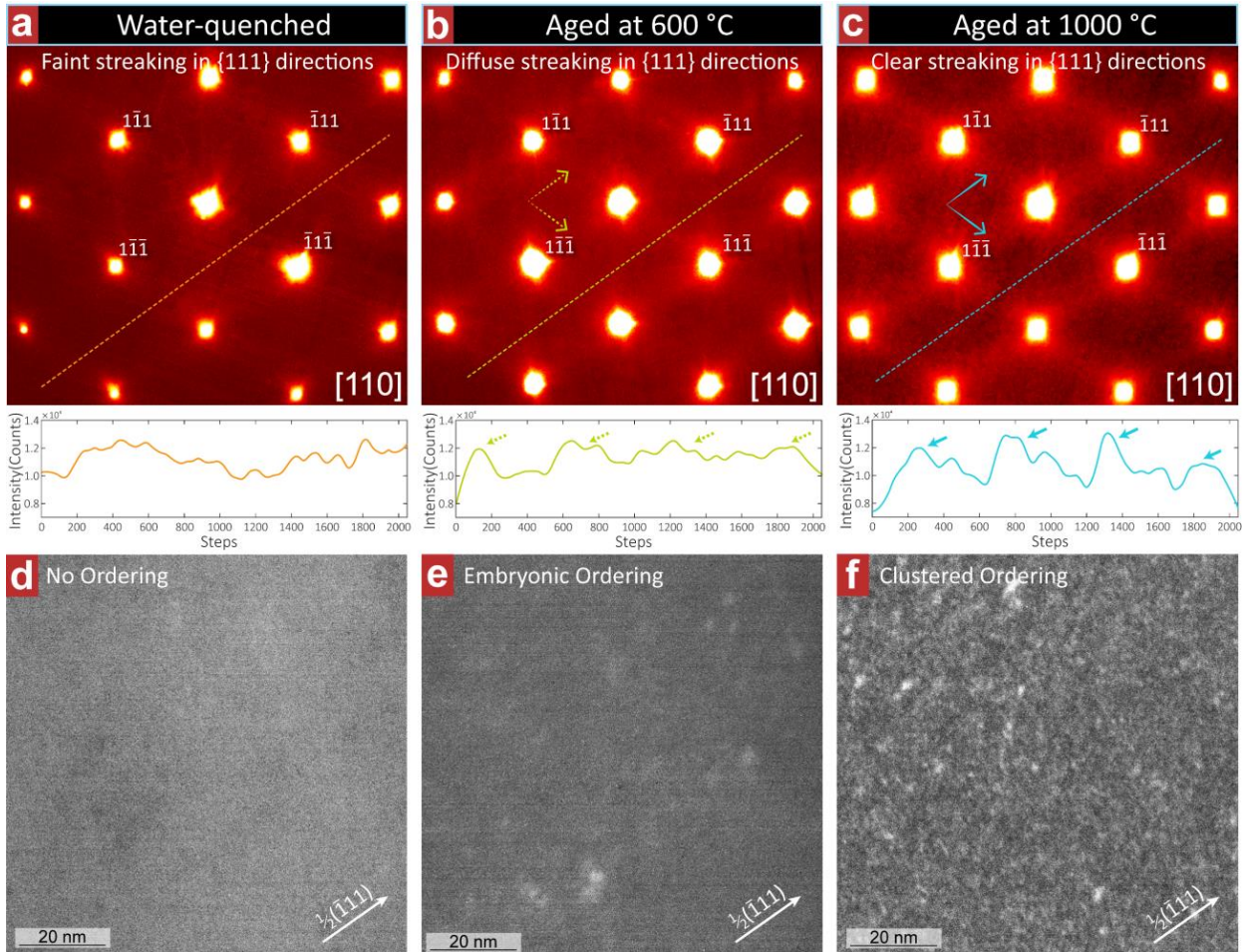

**Supplementary Figure 1:** Energy-filtered TEM diffraction patterns and DF images formed with "diffuse superlattice" streaks. (a-c), energy-filtered diffraction patterns taken from samples that were water-quenched, aged at 600 °C for one week and aged at 1000 °C for one week, respectively. The contrast is reversed and pseudo-colored for better visibility. The line plots of intensity show the periodic intensity of the "diffuse superlattice" streaks. (d-f), energy-filtered DF images taken from water-quenched, 600 °C aged and 1000 °C aged samples, respectively. The aperture positions are marked by the *g* vectors. The images of the water-quenched and the 1000 °C aged samples are the same as in Figure 1 but are presented again here for comparison with the 600 °C aged sample.

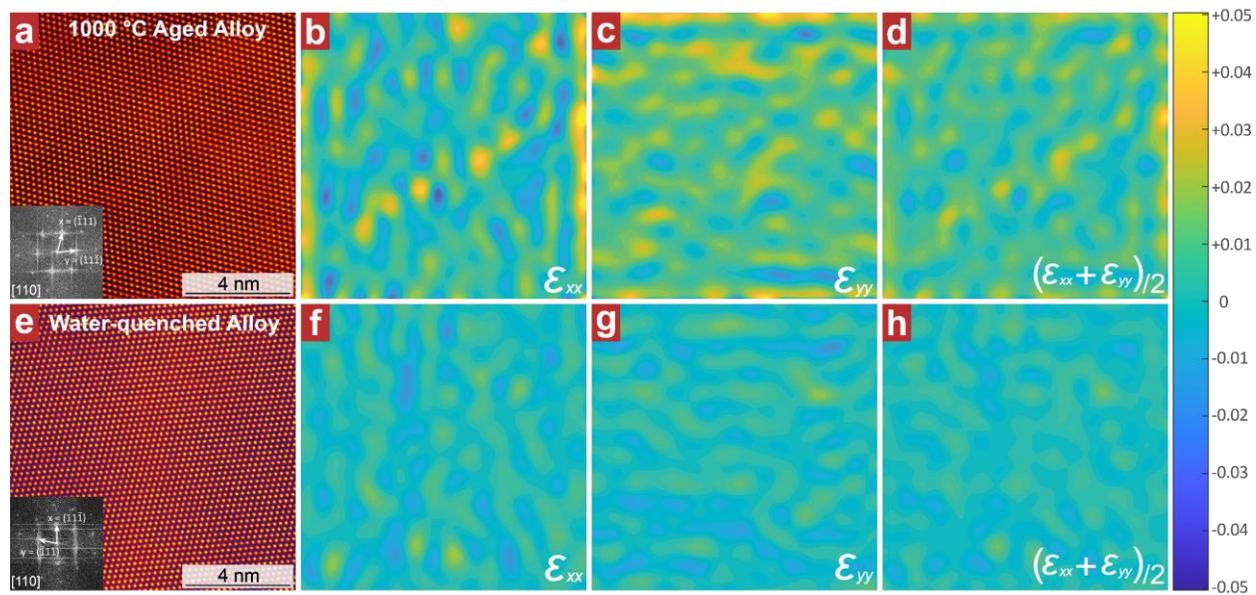

**Supplementary Figure 2:** Geometrical phase analysis strain mapping of a 1000 °C aged sample and water-quenched sample. (a), (e), drift-corrected high-resolution STEM images of the 1000 °C aged sample and the water-quenched sample, respectively. (b) - (d), strain maps of image (a) showing nanometer-sized local fluctuation of strain. (f) - (h), strain maps of image (e) showing similar but much weaker contrast of local strain.

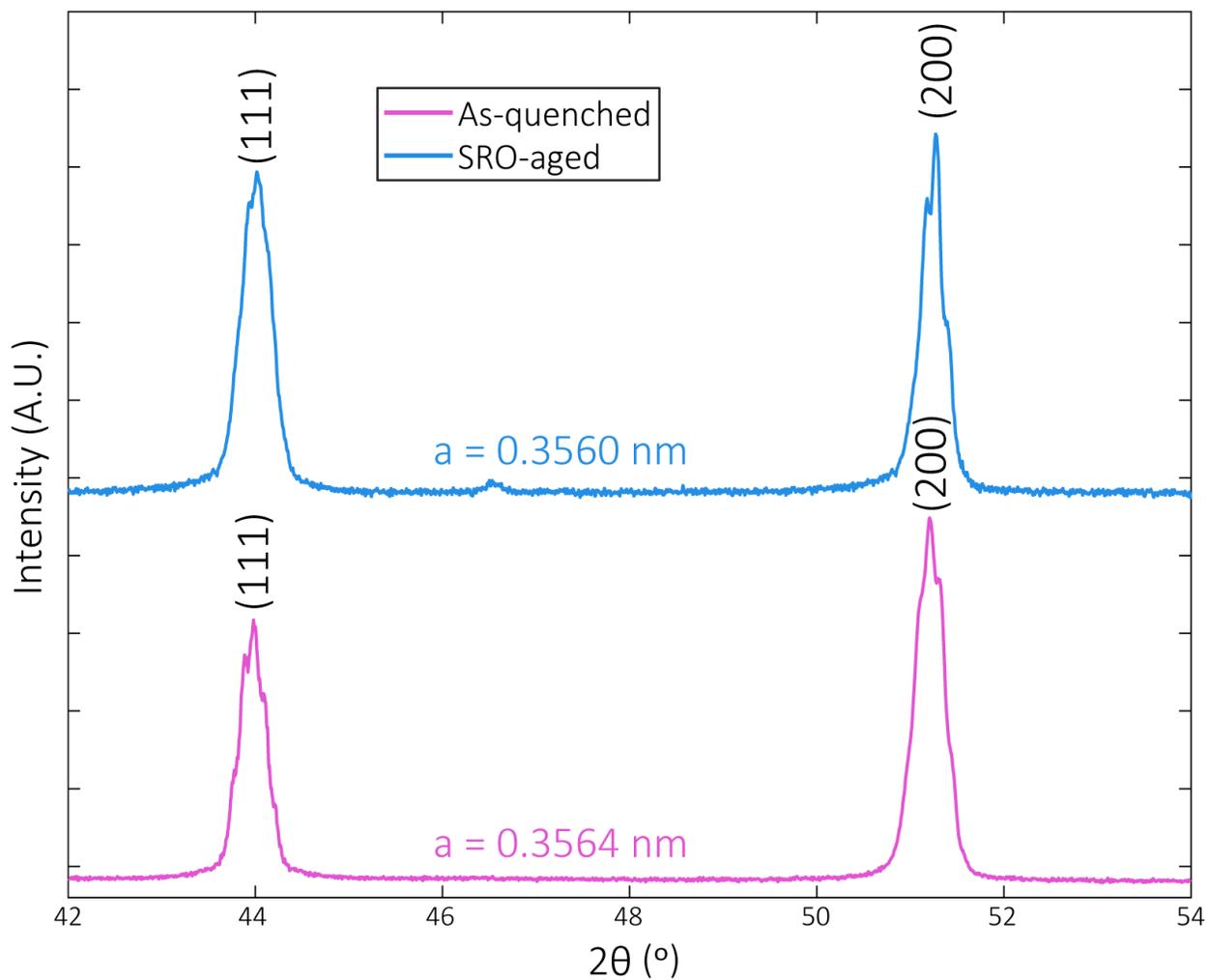

**Supplementary Figure 3:** Results of X-ray diffraction experiments from a water-quenched sample and a 1000 °C aged sample, respectively.

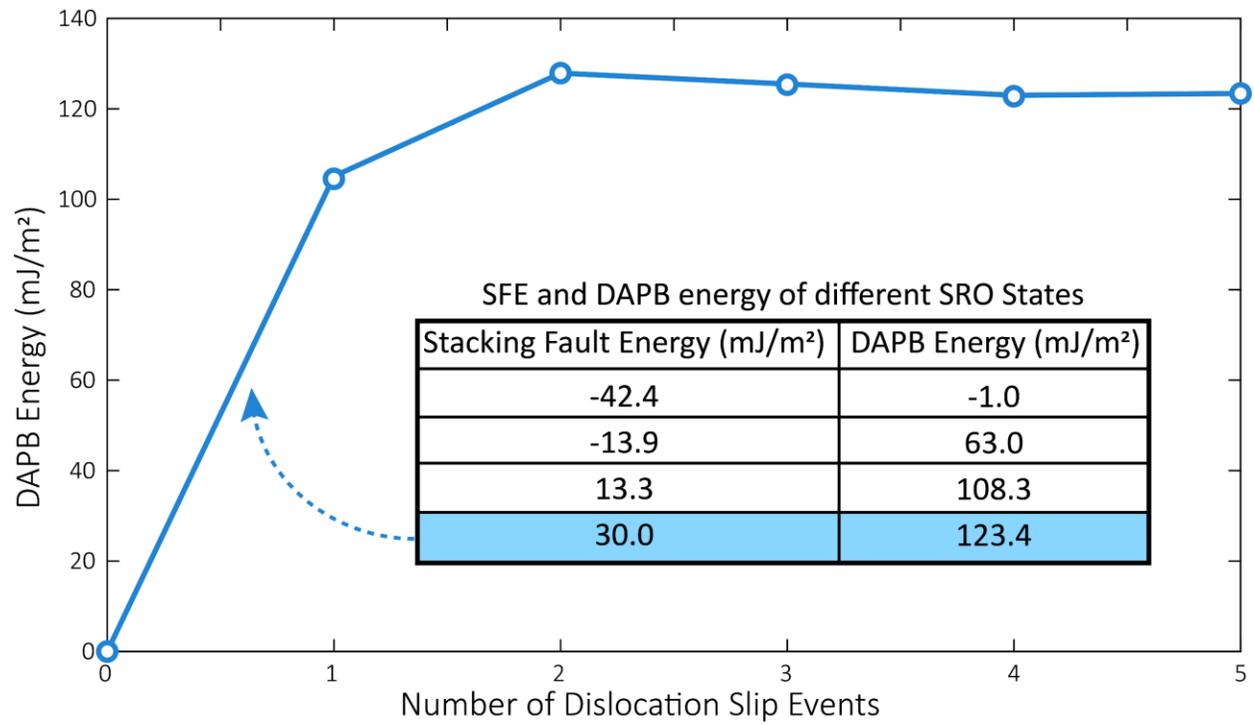

**Supplementary Figure 4:** Diffuse anti-phase boundary energy as a function of successive dislocation slip events from a calculated SRO model. The data in the table represents different states of SRO and the plot is from the state marked blue.

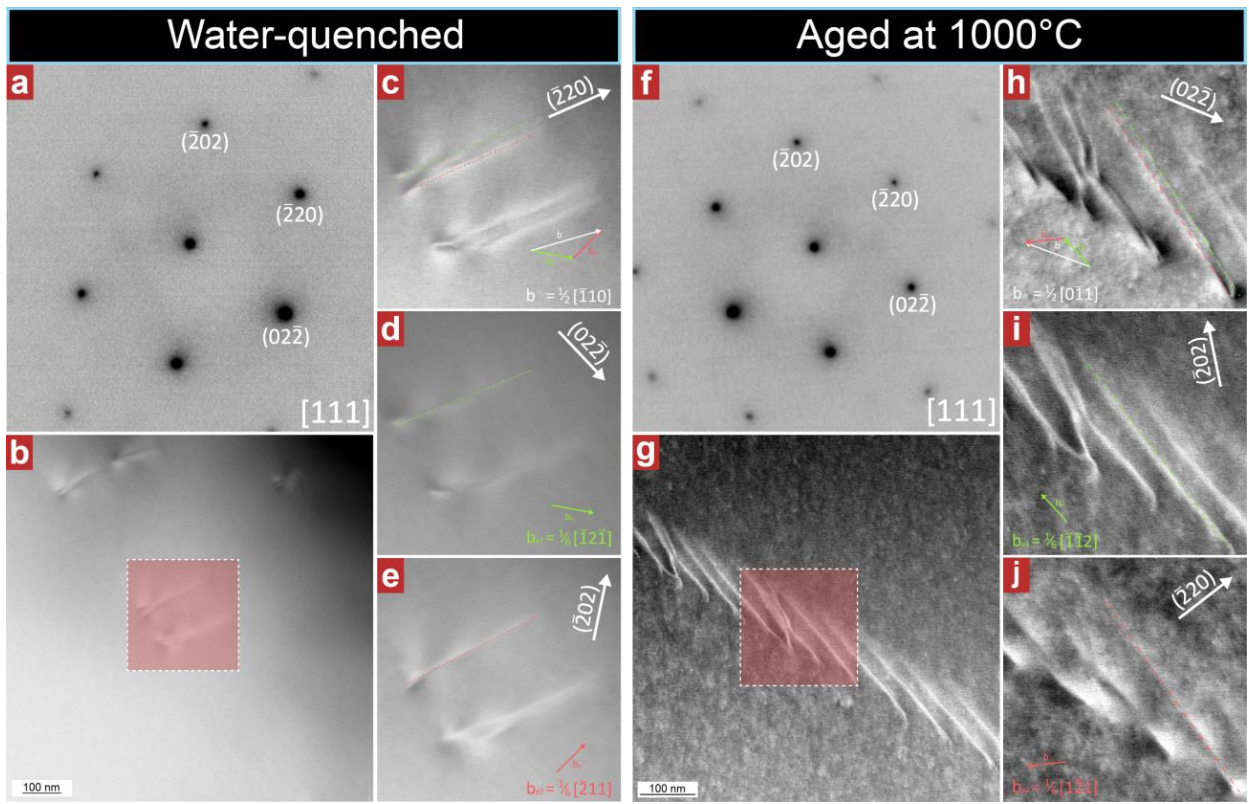

**Supplementary Figure 5:** Detailed "$g \cdot b$" analysis of partial dislocations for the water-quenched (a-e) and aged MEA samples (f-j). (a) and (f) are diffraction references showing the diffraction conditions ($g$ vectors) used for the analysis. (b) and (g) are DC-STEM images showing lower magnification images of dislocations in the water-quenched and aged samples, respectively. (c-e) and (h-j) are two-beam DC-STEM images with the Burgers vectors of the visible dislocations noted on the images.

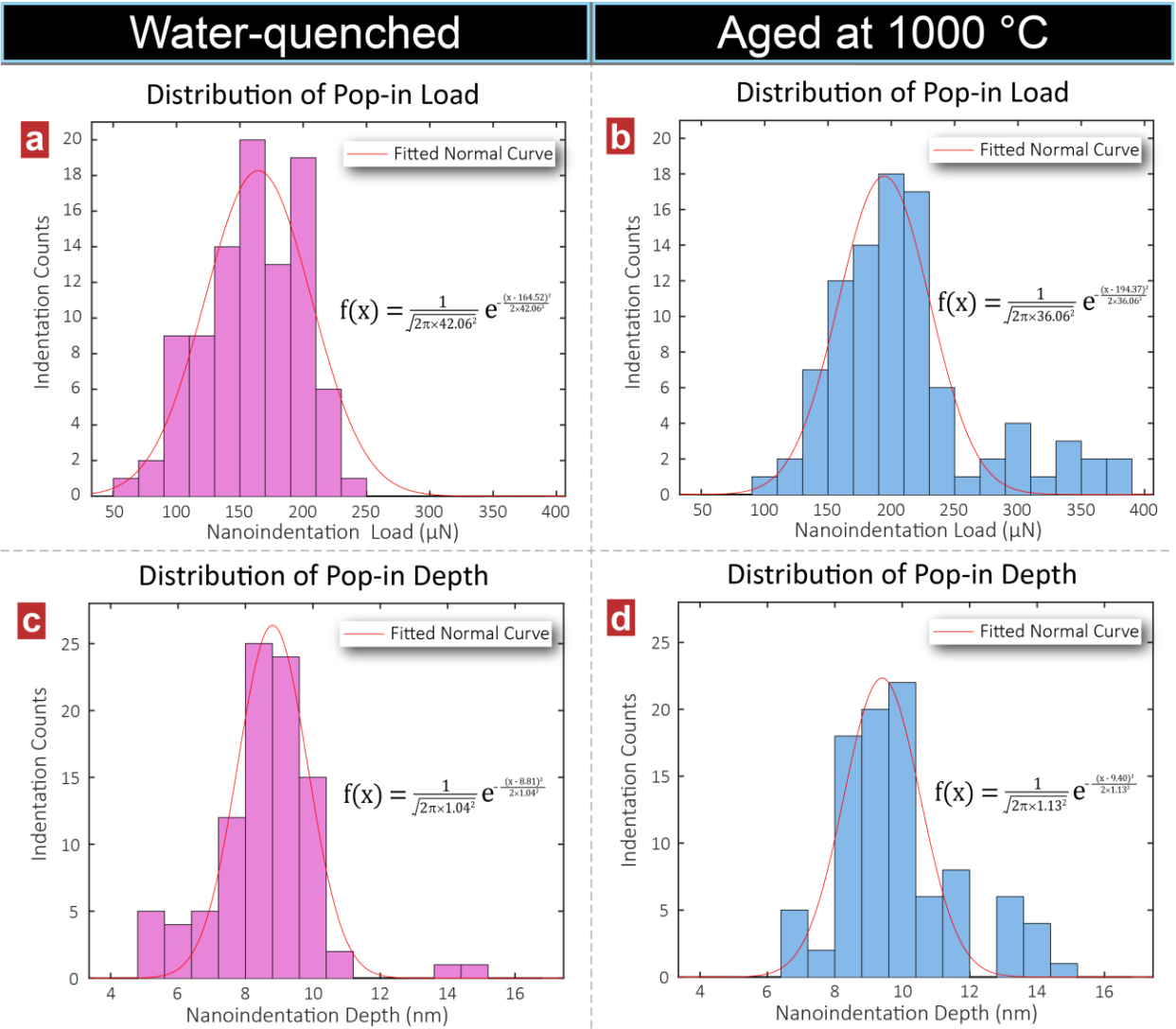

**Supplementary Figure 6:** Detailed statistical analysis of the pop-in events. (a), (b), distribution of the pop-in load from water-quenched and 1000 °C aged samples, respectively. (c), (d), distribution of the pop-in depth from water-quenched and 1000 °C aged samples, respectively. The fitted normal distribution functions are listed in the figures. The results of numerical analysis are summarized in Table 1.

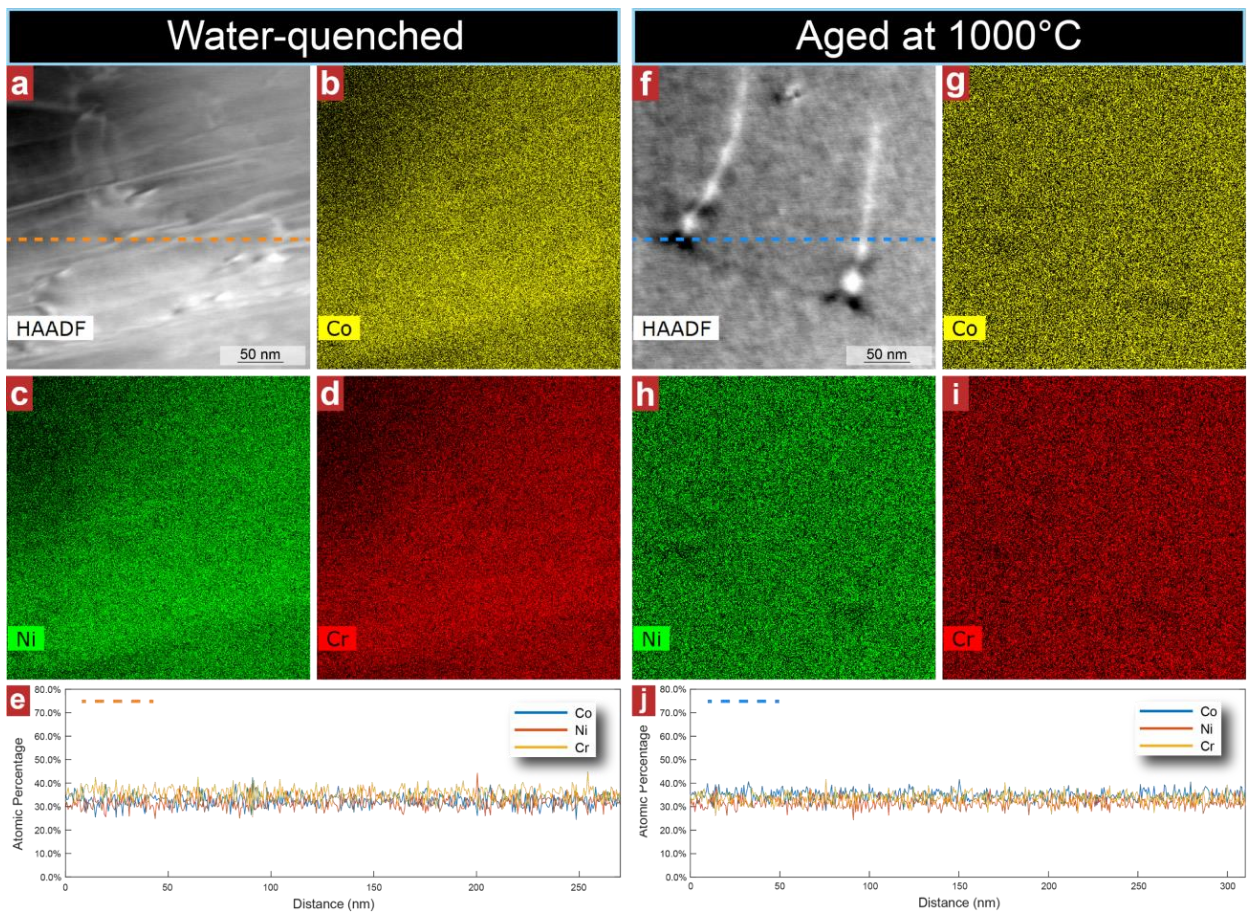

**Supplementary Figure 7:** Results of energy dispersive X-ray mapping (EDS) of the water-quenched and aged MEA samples. (a) and (f) are reference HAADF images showing the regions of interest of a water-quenched sample and a 1000 °C aged sample, respectively. (b) – (d) and (g) – (i) element mapping of Co, Ni and Cr of the water-quenched sample and the 1000 °C aged sample, respectively. (e) and (j) are quantitative results of line scans of the water-quenched sample and the 1000 °C aged sample, respectively. The line scan directions are marked by the dashed lines in (a) and (f).